\documentclass[twocolumn,showpacs,preprintnumbers,amsmath,amssymb,prb]{revtex4}

\usepackage{graphicx}
\usepackage{dcolumn}
\usepackage{bm}
\usepackage{tabularx}
\usepackage{amsmath}

\begin{document}

\title{Reply to ``Comment on `Low-temperature phonon thermal conductivity of single-crystalline Nd$_2$CuO$_4$: Effects of sample size and surface roughness' "}

\author{S. Y. Li,$^1$ J.-B. Bonnemaison,$^1$ A. Payeur,$^1$ P. Fournier,$^{1,2}$ C. H. Wang,$^3$ X. H. Chen,$^3$ and Louis Taillefer$^{1,2,*}$}

\affiliation{$^1$D{\'e}partement de physique and RQMP,
Universit{\'e} de Sherbrooke, Sherbrooke, Canada\\
$^2$Canadian Institute for Advanced Research, Toronto, Canada\\
$^3$Hefei National Laboratory for Physical Science at Microscale and
Department of Physics, University of Science and Technology of
China, Hefei, Anhui 230026, P. R. China}

\date{\today}

\begin{abstract}

In their comment \cite{Sun} on our study of phonon heat transport in
Nd$_2$CuO$_4$,\cite{Li} Sun and Ando estimate that the phonon mean
free path at low temperature is roughly half the width of the single
crystal used in our study, from which they argue that phonon
scattering cannot be dominated by sample boundaries. Here we show
that their use of specific heat data on Nd$_2$CuO$_4$, which
contains a large magnetic contribution at low temperature that is
difficult to reliably extract, leads to an underestimate of the mean
free path by a factor two compared to an estimate based on the
specific heat data of the non-magnetic isostructural analog
Pr$_2$CuO$_4$. This removes the apparent contradiction raised by Sun
and Ando.

\end{abstract}

\pacs{72.15.Eb, 74.72.-h}

\maketitle

Recently, we reported the effect of sample size and surface
roughness on the phonon thermal conductivity $\kappa_p$ of
Nd$_2$CuO$_4$ single crystals down to 50 mK.\cite{Li} At 0.5 K,
$\kappa_p$ was shown to be proportional to $\sqrt{A}$, where $A$ is the
cross-sectional area of the sample. This clearly demonstrates that
$\kappa_p$ is dominated by boundary scattering at and below 0.5 K.

In their Comment,\cite{Sun} Sun and Ando (SA) calculated the phonon
mean free path $l$ from our thermal conductivity data, using the
formula $\kappa_p = C <v> l / 3$ where $C = \beta T^3$ is the phonon
specific heat and $<v>$ is a suitable average of the three acoustic sound
velocities. Using $\beta = 0.42$ mJ/mol K$^4$ for Nd$_2$CuO$_4$,
they arrived at an estimate of $l$ which is about 50\% of the average sample
width $W \equiv 2 \sqrt{A/\pi}$.
Because $l < W$, they concluded that phonons in our Nd$_2$CuO$_4$ single crystals
did not reach the boundary scattering regime at low temperature.

Because their criticism is based entirely on a quantitative estimate
of $l$, it would seem important to know the uncertainty in the
parameters $\beta$ and $<v>$ used to arrive at this estimate. SA
provide no indication of the uncertainty on their numbers. More
problematic, however, is their use of specific heat data on
Nd$_2$CuO$_4$ to estimate $\beta$. Indeed, because of the large
magnetic contribution to $C(T)$ at low temperature coming from
Nd$^{3+}$ moments, SA had to make some assumptions to extract the
phonon component from that low-temperature data. An alternate and
standard approach is to use the non-magnetic isostructural analog
Pr$_2$CuO$_4$, for which there is no such magnetic contribution. The
specific heat of Pr$_2$CuO$_4$ below 10 K readily yields $\beta$ =
0.19 mJ/mol K$^4$.\cite{Ghamaty}

If we use that value for $\beta$, which is half that used by SA for
their estimate of $l$, we arrive at an estimated phonon mean free
path which is twice as large. This removes the discrepancy which is
the basis for their Comment. Note also that $\beta$ = 0.19 mJ/mol
K$^4$ is in excellent agreement with the value of $\beta$ = 0.20
mJ/mol K$^4$ calculated by SA using the standard expression based on
sound velocities (Eq.(2) in their Comment).

We conclude that as best as one can estimate it, the phonon mean
free path in Nd$_2$CuO$_4$ at $T = 0.5$~K works out to be roughly
equal to the average sample width, so that one can expect phonons to
be scattered by boundaries below 0.5 K. In as-grown crystals with
smooth, mirror-like faces, it is then reasonable to expect some
specular reflection as the phonon wavelength gets longer and longer
with decreasing temperature. The best way to verify this is to
deliberately roughen those surfaces and see whether the thermal
conductivity is reduced. The purpose of our article\cite{Li} was to
show that it clearly does.

$^*$ E-mail: louis.taillefer@physique.usherbrooke.ca


\begin{thebibliography}{99}

\bibitem{Li} S. Y. Li, J.-B. Bonnemaison, A. Payeur, P. Fournier, C. H. Wang, X. H. Chen, and Louis Taillefer, Phys. Rev. B {\bf
77}, 134501 (2008).
\bibitem{Sun} X. F. Sun and Yoichi Ando, arXiv:0904.1704.
\bibitem{Ghamaty} S. Ghamaty, B. W. Lee, J. T. Markert, E. A. Early, T. Bjornholm, C. L. Seaman, and M. B. Maple, Physica C {\bf
160}, 217 (1989).


\end{thebibliography}
\end{document}